# AIomics: exploring more of the proteome using mass spectral libraries extended by AI

Lewis Y. Geer[1*], Joel Lapin[2,3], Douglas J. Slotta[1], Tytus D. Mak[1], Stephen E. Stein[1]

[1]*Mass Spectrometry Data Center, National Institute of Standards and Technology, Biomolecular Measurement Division, 100 Bureau Dr., Gaithersburg, Maryland 20899, United States*

[2]*Department of Physics, Georgetown University, Washington, DC 20057, United States.*

[3]*Associate, Mass Spectrometry Data Center, National Institute of Standards and Technology, Biomolecular Measurement Division, 100 Bureau Dr., Gaithersburg, Maryland 20899, United States*

[*]Email: lewis.geer@nist.gov

## Abstract

The unbounded permutations of biological molecules, including proteins and their constituent peptides, presents a dilemma in identifying the components of complex biosamples. Sequence search algorithms used to identify peptide spectra can be expanded to cover larger classes of molecules, including more modifications, isoforms, and atypical cleavage, but at the cost of false positives or false negatives due to the simplified spectra they compute from sequence records. Spectral library searching can help solve this issue by precisely matching experimental spectra to library spectra with excellent sensitivity and specificity. However, compiling spectral libraries that span entire proteomes is pragmatically difficult. Neural networks that predict complete spectra containing a full range of annotated and unannotated ions can be used to replace these simplified spectra with libraries of fully predicted spectra, including modified peptides. Using such a network, we created predicted spectral libraries that were used to rescore matches from a sequence search done over a large search space, including a large number of modifications. Rescoring improved the separation of true and false hits by 82%, yielding an 8% increase in peptide identifications, including a 21% increase in nonspecifically cleaved peptides and a 17% increase in phosphopeptides.

## Introduction

Proteomics is the large-scale study of the entire set of proteins that are found in a cell, a tissue, or an organism. This comprehensive scope presents unique challenges to the experimentalist. While a genome may contain tens of thousands of genes, the number of structurally different proteins translated from the genome exponentially increases through variations such as isoforms, mutations and post-translational modifications (PTMs). Commonly, these proteins are identified by enzymatically digesting them into their constituent peptides, injecting them into a mass spectrometer, and dissociating them, with the resulting fragment mass/charge and intensities measured to generate peptide mass spectra. Sequence search algorithms[1–7] use computationally digested protein sequence records to generate theoretical peptide spectra to match to the experimental spectra. These theoretical spectra contain the mass/charge values of a limited number of possible ion types from canonical dissociation reactions, and typically the algorithms do not attempt to predict whether these ions will be realized or, if they are, their intensities. These factors can reduce the specificity of search results, particularly when searching large chemical spaces. To improve specificity and to minimize search time, the sequence search algorithms significantly limit the types of peptides being searched and modifications considered, such as requiring digestion to occur perfectly.

Due to these limitations, sequence searches are subject to the streetlight effect: they only search within a limited space, a significant constraint when looking for novel biological results. To address this issue, open searches[8–15] algorithmically modify sequence searches to allow for a much larger breadth of variations in the chemical structure of proteins. However, they typically require an unmodified spectrum of the peptide being examined and the larger space searched comes with a consequence: false positive matches to the theoretical spectra of molecules that are misleading or chemically inexplicable[16].

Spectral library searching[12,17–20] holds at least a partial solution to this issue by matching the observed spectra to those from a library of previously identified experimental spectra. Spectral libraries, unlike the theoretical spectra generated for sequence search algorithms, contain only the ion mass/charge and intensities that are the experimentally observed products of fragmentation reactions. Due to these factors, spectral library searching has been shown to have better performance than sequence search algorithms[21], but can suffer from incomplete proteome coverage since spectral libraries must be compiled and curated from existing sets of experimental spectra which may not span all known protein sequences and modifications.

To increase the proteome coverage of spectral libraries, it is possible to predict mass spectra from a peptide sequence using computational models of peptide fragmentation[22,23]. One such model, MassAnalyzer[24,25], estimates the equilibrium constants for fragmentation reactions. However, libraries generated from MassAnalyzer predictions have not consistently produced improvements in search results[21].

Approaches using AI, in contrast to an explicit model of fragmentation reactions, use a general model architecture to learn how to predict spectra by training directly on large sets of spectral data[26–31]. In particular, deep neural networks (DNN) have reported significant progress in predicting spectra. The DNN Prosit[32,33] directly predicts the abundances of only b and y ion types. This approach is not able to predict unannotated ions and requires a dataset where ions are well annotated. A similar DNN, pDeep[34–36], predicts a limited set of fragmentation ions while allowing for 21 modifications, including phosphorylation which plays a prominent role in biology, including in cellular signaling[37,38]. One intriguing class of networks predicts full spectra that, in contrast to previous networks, do not require annotated training spectra and predict complete spectra that include unannotated ions. Not requiring annotations not only makes training networks easier, but also allows the network to be trained for any set of modifications. PredFull[39] is a seminal DNN in this class. In comparison to previous approaches, we use a transformer-based network[40,41] that predicts the complete spectra of modified and unmodified peptides that are dissociated using Higher Energy Collisional Dissociation (HCD). The network was trained with the diverse, high-quality spectra in the NIST peptide spectral library[42,43]. The ability to predict unannotated ions may be important for HCD spectra, where 44% of all ions can be unannotated[44].

The use of DNN spectrum prediction networks to improve search results is a subject of active investigation. The INFERYS[45] algorithm was shown to boost peptide identifications from the Sequest HT sequence search algorithm using a DNN that predicts spectra with a limited set of ion series. The Prosit[46,47] approach based on the same DNN architecture also shows boosting of search results. Both INFERYS and Prosit are largely limited to a small set of modifications that typically form during sample preparation. The Deephos[48] search strategy uses a network that predicts neutral loss ions and a novel search algorithm to improve the identification of only TMT-labeled phosphopeptides, although it is not clear if the improvements shown are due to the predictions or to features of the search algorithm, including its reduced search space.

Because the network we used predicts the complete spectra of peptides that are typically found in spectral libraries, including those with PTMs and modifications like TMT, we set out to understand not only if the use of DNNs boosted the search results of typical peptide classes, but also whether it would allow us to significantly expand the chemical space being searched. To do this, we performed an exhaustive sequence search over a much larger chemical space of atypically cleaved semi-tryptic peptides decorated with a large number of modifications, where semi-tryptic peptides are those cleaved at one end by trypsin at the C-terminal side of arginine and lysine but not at any specific site on the other end. The results of the search were rescored using a predicted spectral library, and a False Discovery Rate (FDR) analysis was done on the results. Rescoring allows us to use the sequence search results as a control, isolating the effects of what is effectively a spectral library search on predicted spectra. This technique enables the extrapolation of our conclusions to the general use of spectral library searching with predicted spectra.

# Methods

## Data

We used NIST high-resolution mass spectral libraries[49], created using Orbitrap HCD spectra from human, mouse and Chinese hamster ovary cells (cell lines CHO-K1, CHO-S, CHO-DG44). The only libraries omitted were from iTRAQ experiments. The construction of the NIST spectral libraries has been discussed previously[42,50]. These libraries contain both consensus spectra and "best of" spectra from a set of replicates. The libraries include several modifications: carbamidomethyl cysteine, methionine oxidation, glutamine to pyro-glutamic acid, glutamic acid to pyro-glutamic acid, N-terminal acetylation, N-terminal pyro-carbamidomethyl cysteine, serine/threonine/tyrosine phosphorylation, and TMT6plex. Phosphopeptide identifications were taken from Sharma *et al.*[51]

The creation of test and validation sets had an additional step beyond random selection to ensure that their peptides were dissimilar from those in the training set. First, subsets of the complete non-redundant list of peptides were randomly chosen to be in either the validation or test sets. Peptides that mapped only to consensus spectra were moved to the training set. Next, all of the peptides in the validation set and test set were compared against each other and against the peptides in the training set using a normalized Levenshtein score to ensure their dissimilarity. Peptides in the validation and test sets that were too similar to peptides in the training set were removed from the validation and test sets and placed into the training set. Peptides were considered similar if their score was greater than 0.7 using the normalized similarity measure, which ranges from 0 for two strings requiring the maximum number of possible edits to 1.0, which denotes an exact string match. For purposes of testing, consensus spectra were filtered from the test and validation sets after their creation.

The network's input embedding takes a peptide of length up to 40, charges 1+ to 8+, normalized collision energy (NCE), and the previously mentioned modifications. As is typical of proteomics data, the large majority of training spectra fall in a narrower range of characteristics than the extent of the data would suggest. For the training dataset, 96% of the spectra have charge states from 2+ to 4+, NCEs ranging from 25% to 38%, and peptide length from 6 to 30 residues. For the test and validation sets, only 94% of the spectra meet those criteria. The statistics of the datasets by the total number of spectra, mean peptide length, and the percentage of phosphopeptide spectra are shown in Table 1. The requirement that the test and validation sets contain sequence dissimilar peptides tends to enrich for longer peptides. The phosphopeptide spectra in the NIST library are not consensus spectra, so there are higher percentages of them in the test and validation sets.

**Table 1. Dataset statistics[a]**

|             | training | validation | test  |
|-------------|----------|------------|-------|
| **all**         | 2295094  | 10684      | 71484 |
| **non-tryptic** | 161357   | 331        | 2908  |
| **phosphorylated** | 57268 | 743 | 4449 |
| **TMT** | 569200 | 1852 | 13263 |
| **oxidized** | 260474 | 1765 | 10790 |
| **acetylated** | 11170 | 40 | 201 |
| **pyro-glutamic acid** | 3954 | 7 | 66 |
| **pyro-CMC** | 861 | 1 | 22 |
| **peptide length** | 14 | 19 | 18 |

[a] summary statistics for the training, validation, and test data sets. Numbers of total spectra, non-tryptic spectra and spectra with various modifications are summarized, followed by the mean peptide length for each dataset.

---

## Software Libraries

The evaluation methods and DNN were coded in Python using the PyTorch[52], Bayesian-Torch[53], Lightning[54], Arrow, SciPy[55], Pandas[56], and Numpy[57] libraries. Networks were trained on NVIDIA DGX-1 nodes. The evaluation and network code can be found at https://github.com/usnistgov/masskit and https://github.com/usnistgov/masskit_ai. The network contains 21M parameters. While use of the network is more efficient using a GPU, our libraries can run under a CPU as detailed in the documentation.

## Losses and Metrics

The network takes the peptide sequence, precursor charge, modifications, and normalized collision energy as input. It then outputs a predicted spectra as an array of intensities indexed by mass/charge with fixed bin sizes of 0.1 Da as we found that smaller bin sizes lengthened training and did not improve scores. The loss used for training the network is a standard cosine similarity $C$ applied to the predicted and library spectra expressed as fixed bin arrays. The evaluation metric used is the Stein-Scott[58] dot product

$$S = 999 \frac{\left( \sum_{(i,j) \in M} \sqrt{Q_i L_j} \right)^2}{\sum_i Q_i \sum_j L_j} \quad (1)$$

where $Q_i$ are query spectrum ion intensities and $L_j$ are library spectrum ion intensities. $M$ is the set of all matched ions between the query and library spectra. Per convention, $S$ ranges from 0-999, where 999 indicates a perfect match. When used as a similarity metric for evaluation, $S$ is calculated using interval arithmetic where each interval corresponds to the peak width of each ion in each spectrum. Interval arithmetic is used for evaluation as it more precisely represents ppm resolution and allows for peaks of 0 width generated during upscaling, a procedure discussed below.

## Annotating and Upscaling Predicted Spectra

To evaluate the ability of the network to predict various ion types, we generated theoretical spectra using the corresponding peptide sequence, precursor charge, and modifications for each test experimental spectrum. These theoretical spectra consisted of immonium ions, internal ions, parent ions, $b^{1+}$, $y^{1+}$, $y^{2+}$, $y^{3+}$, and $a^{1+}$ ions. Neutral losses of water and ammonia were annotated for ion series of charge 1 and the parent ions. For phosphopeptides, the neutral losses for the ion series included $H_3PO_4$ and $HPO_3$. These theoretical ions were matched to the experimental spectra. Predicted spectra were then generated and matched to the experimental spectra and any annotations on the experimental ions were applied to the matching predicted ions. If multiple experimental ions matched a predicted ion, the experimental ion with the closest intensity to the predicted ion was selected as a tiebreaker.

*Upscaling*. This annotation mechanism also allows us to improve the mass/charge resolution of many predicted ions by first matching predicted spectra directly to theoretical spectra. When a match between a theoretical and predicted ion is unambiguous, the mass/charge of the theoretical ion replaces the mass/charge of the predicted ion, yielding a resolution that is equal to that of the theoretical ion calculations, +/- 0 ppm. 80% of the predicted ions with intensity > 5% of maximum were upscaled in our test set.

## Sequence Search Evaluation

The test set, minus spectra with mass > 4200 daltons, were searched with Mascot 2.6[2] against the UniProt[59] Chinese hamster, mouse and human proteomes, including isoforms. Decoy databases were created using OpenMS[60] DecoyDatabase. Search settings different from the default were: semi-tryptic cleavage, 2 missed cleavages allowed, 20 ppm precursor ion tolerance, 0.02 Da product ion tolerance, 1 $^{13}$C allowed, ESI-QUAD-type fragmentation, fixed modification of carbamidomethyl (C) and variable modifications of oxidation (M), acetyl (protein N-term), phospho (ST), phospho (Y), gln->pyro-glu, glu->pyro-glu, TMT6plex (K), TMT6plex (N-term), and pyro-carbamidomethyl.

For each query spectrum, we deleted any decoy matches that were the reverse of the query sequence, including and not including the C-terminal residue of both peptide sequences. For each peptide/spectrum match (PSM), the network was used to predict a spectrum for each matched peptide, using the same NCE as the query. A corrected $S$ score was calculated between each predicted spectrum and the corresponding query spectrum. As shown in Figure S1, a correction was required as the significant threshold of the $S$ score is dependent on the number of matched ions.

The 1% FDR score thresholds for the Mascot ions score and the corrected $S$ score are calculated assuming matches with identical peptide sequence, charge, modifications and modification positions are true matches ($T$) and decoy matches are false matches ($F$). The FDR was determined by effectively searching the original and decoy databases separately and taking the top PSMs above threshold. To avoid overestimating the FDR, we omitted false hits to the original sequences. The expression used to calculate the FDR was $F/(T+F)$[61].

# Results and Discussion

## Prediction Performance

**Figure 1**

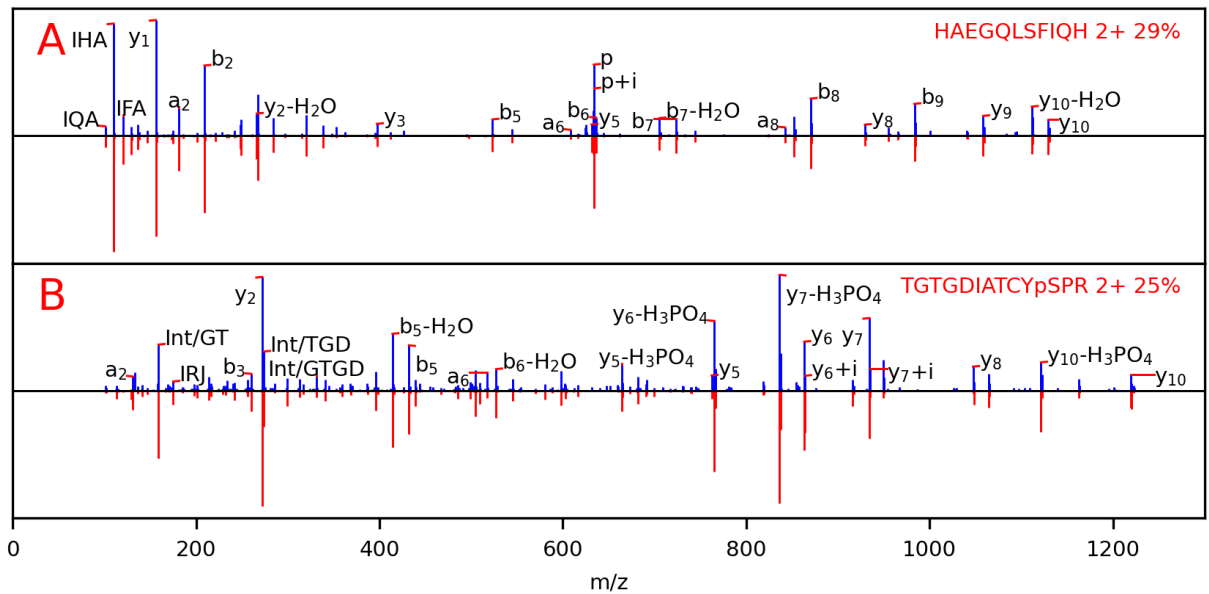

Figure 1. Mirror plots of (A) a non-tryptic spectrum with score near the median prediction score and (B) a phosphopeptide spectrum with score near the bottom 10th percentile score. The blue spectrum at the top of each mirror plot is the experimental spectrum from the test set, annotated by ion series, including immonium ions (IQA, etc.), parent ions (p), and ions containing carbon-13 (+i). The red spectrum below is the matching predicted spectrum. The predicted phosphopeptide spectrum contains neutral loss ions that are useful for identifying and localizing phosphosites, as well as internal ions (Int/), immonium ions, and unannotated ions.

Using the DNN, we predicted spectra that correspond to those from our test set, yielding a median similarity $S$ of 695. Figure 1A displays an example non-tryptic peptide spectrum at the median score, where almost all major ions have been predicted with largely accurate intensities. Figure 1B is a representative phosphopeptide spectrum scoring at the test set's 10th percentile of $S$ scores, 536, showing that the majority of major ions are well predicted. Figure 2A is a histogram of all test set $S$ scores. The network accurately predicts a wide variety of ion types, including ions generated from unknown mechanisms, that are from peptides dissimilar to those found in the training set. When comparing score values to other studies, it is important to note that, unlike some previous studies, the test set is sequence dissimilar to the training set.

Additionally, all ions are included when scoring (not just an abridged set) and the score $S$ is computed on the square roots of the intensities to avoid over-emphasizing high intensity ions. As detailed in Figures S2-S4, the results on the test set indicate that the network can make useful predictions on a wide range of peptide sizes, charges, and NCE values. Spectra with poor scores tend to have long amino acid repeats.

**Figure 2**

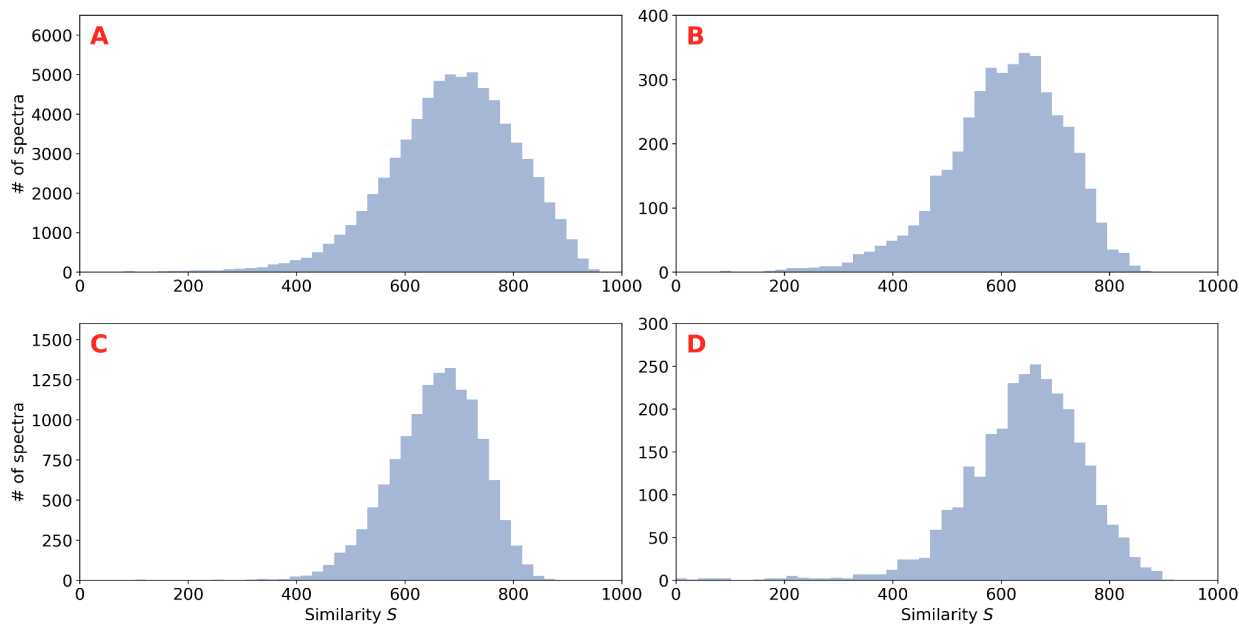

Figure 2. (A) Histogram of the similarity score $S$ calculated between the experimental spectra and predicted spectra in the test set. (B) Histogram of $S$ for the subset of the test set that contains the spectra of phosphopeptides. (C) Histogram of $S$ for TMT derivatized peptides. (D) Histogram of $S$ for non-tryptic peptides.

Of particular biological interest is the identification and localization of phosphorylation PTMs using spectra. A histogram of the similarity score $S$ for phosphopeptides in the test set is shown in Figure 2B. The median score for phosphopeptide predictions is 617, with 90% of the predicted spectra scoring above 468. This indicates that the network predicts phosphopeptide spectra slightly worse than non-phosphopeptide spectra, which may be due to the fact that the training set contains only approximately 57,000 phosphopeptide spectra.

Peptide derivatization by TMT is a widely used technique in quantitative proteomics. The test set contains a number of TMT derivatized peptides, and a histogram of the similarity score $S$ of these spectra is shown in Figure 2C. The predicted TMT spectra have a median score of 662 with 90% of the spectra scoring above 547, indicating that the prediction network may be useful in experiments that quantify protein abundance using TMT.

Non-tryptic peptides do not have arginine or lysine at their C-terminus. These peptides, found in our test set at the C-terminus of proteins or due to imperfect cleavage by trypsin, mimic nonspecifically cleaved peptides found in nature, including in the immunopeptidome. Figure 2D is a histogram of the $S$ score for these peptides. The spectra predicted for these peptides have a median $S$ score of 654 where 90% of the predictions score above 512, indicating the usefulness of the model for real world experiments where enzymes do not cleave perfectly, when other enzymes are used for protein digestion, and when analyzing naturally occurring peptides, such as those presented as antigens to the immune system.

## Using Spectral Predictions to Rescore Sequence Search Results by Spectral Library Matching

Since we have found that the predicted spectra can closely model real spectra, we next set out to see if the predicted spectra could be used to improve the number of identified spectra when searching a large chemical space consisting of semi-tryptic peptides, multiple possible modifications, and multiple organisms. The search was performed using a popular sequence search algorithm, Mascot, and rescored using the same procedure as a spectral library search. As described in the methods, we searched the test set against the human, mouse and Chinese hamster proteomes. We also searched against reversed versions of these proteomes, which served as a decoy set to calculate the FDR rate of the searches. After combining all of the top PSMs from these searches, we predicted spectra for each match, using the NCE of the query spectrum when generating the predictions. For each PSM we calculated a corrected $S$ score between the predicted spectrum and the query spectrum. The correction, detailed in the Supporting Information, adjusts the similarity score $S$ by subtraction so that the score threshold is independent of the number of ions, and the correction can be negative to indicate lack of significance. Each PSM where the peptide sequence, charge, modifications, and modification locations of the match were identical to the query was designated a true match and all other matches were designated false.

Figure 3A depicts the histogram of the Mascot ions score for both true matches and false matches. Figure 3B shows the same histogram for the corrected $S$ score. Decoy matches are not included in these histograms although they have the same general distribution as the false matches. The corrected $S$ score performs several times better than the ions score in separating true matches from false matches. To compute this difference, we fit normal distributions to each true and false peak using maximum likelihood estimation, and then calculated the resolution as $(\mu_t - \mu_f)/(\sigma_t + \sigma_f)$ where $\mu_t$ and $\sigma_t$ are the mean and standard deviation of the true peak and $\mu_f$ and $\sigma_f$ are the mean and standard deviation of the false peak. The improvement in resolution was 82% after rescoring.

To understand the effect of unannotated ions in the matches to predicted spectra, we computed corrected $S$ scores using predicted spectra filtered of all unannotated ions. This resulted in an increase in resolution of only 24%, indicating the importance of these unannotated ions. This lower separation can be seen in Figure S5, which is a histogram of the corrected *S* score for predicted spectra that contain only annotated ions. Inspection of predicted spectra that score poorly after filtering reveal spectra with intense unannotated peaks and low intensity annotated peaks.

**Figure 3**

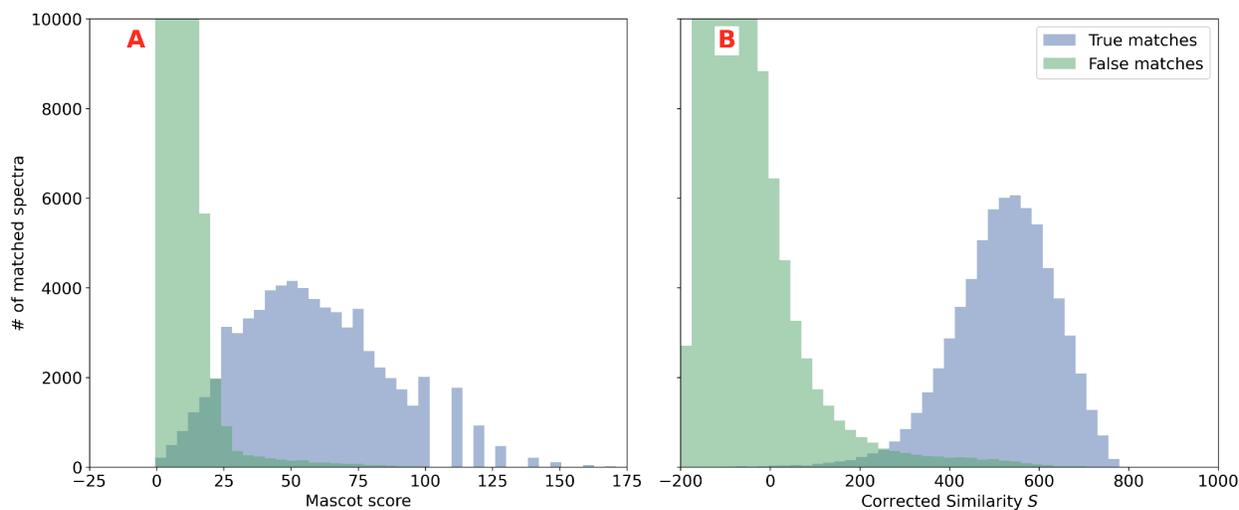

Figure 3. (A) Histogram of the Mascot ions score for both true and false matches to the test set spectra, searching against the human, mouse, and Chinese hamster proteome. (B) histogram of the corrected *S* score as applied to predicted spectra for the same search results as (A). The separation between true and false matches is improved by 82%.

We then performed an FDR analysis of the corrected *S* score and ions score. To accomplish this, we swept the score threshold for both scores and calculated the FDR by using the top true match and top decoy match above the threshold for each test spectrum, taken from all matches returned by Mascot of the test spectra to the three proteomes and their decoys. The results are shown in Figure 4. The number of true matches above threshold for the corrected *S* score exceeds the true matches for the ions score except at very small values of the FDR.

**Figure 4**

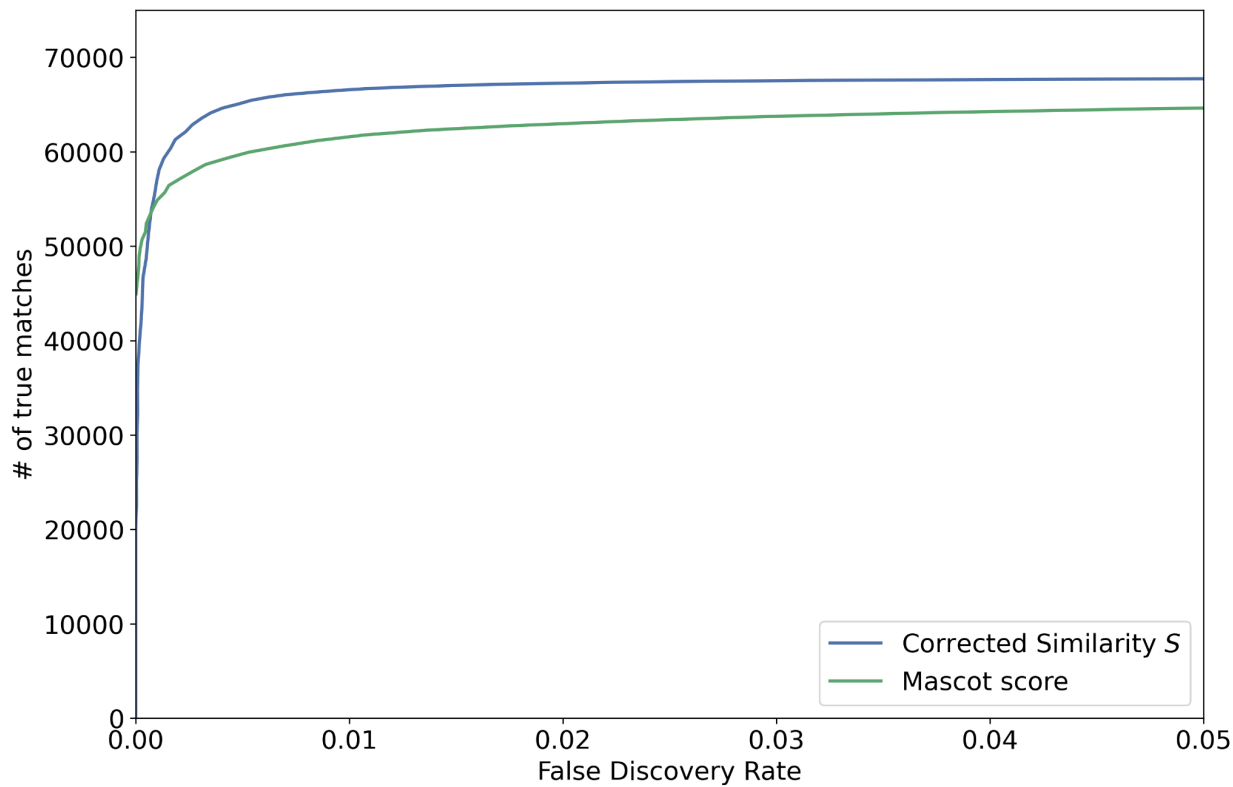

Figure 4. FDR analysis done before and after rescoring using test spectra as the queries. True matches are from the human, mouse, and Chinese hamster proteome and defined as those that match the peptide sequence, charge, and modifications of the query. Spectra were predicted for each Mascot sequence search match and a corrected Stein-Scott dot product $S$ calculated to rescore the matches.

**Table 2: Numbers of peptides identified before and after rescoring at 1% FDR[a]**

|  | both | rescore only | % increase | search only | % decrease |
|---|---|---|---|---|---|
| all | 60321 | 6337 | 10.3 | 1093 | 1.8 |
| unphosphorylated | 58010 | 5883 | 10.0 | 1037 | 1.8 |
| phosphorylated | 2311 | 454 | 19.2 | 56 | 2.4 |
| TMT | 11333 | 1114 | 9.8 | 68 | 0.6 |
| non-tryptic | 2210 | 509 | 22.7 | 35 | 1.6 |

[a] "both" means true positive for both the sequence search and the rescoring, "rescore only" means true positive only after rescoring, and "search only" means true positive only before rescoring.

At 1% FDR, the ions score cutoff is 23 and the corrected $S$ score threshold is 263. Table 2 describes increases and decreases in identifications of true spectra for various subsets of peptides. For unphosphorylated spectra, the net result of rescoring is to gain 8.2% more identifications. Examining phosphopeptides, 16.8% more identifications are made in net after rescoring, losing 2.4% of the search only identifications. TMT derivatized and non-tryptic peptides have a net increase of 9.2% and 20.9%, respectively.

To investigate the causes of these gains, we examined the types of ion series present in true matches before and after rescoring. Unphosphorylated and phosphorylated peptides were examined separately in part due to the importance of neutral losses in identifying phosphopeptides.

**Table 3. Mean Counts of Ion Types in True Matches for Search Results Before and After Rescoring[a]**

|  | unphosphorylated | | | phosphorylated | | |
|---|---|---|---|---|---|---|
|  | both | rescore only | search only | both | rescore only | search only |
| peptide length | 18.2 | 18.4 | 21.0 | 20.7 | 21.9 | 28.1 |
| # ions | 61.4 | 63.0 | 61.7 | 81.5 | 89.1 | 81.5 |
| unannotated | 22.0 | 32.1 | 30.1 | 30.3 | 42.0 | 36.7 |
| y ions | 18.5 | 14.8 | 15.0 | 26.6 | 26.2 | 24.2 |
| b ions | 8.9 | 5.9 | 6.2 | 11.7 | 8.0 | 10.1 |
| a ions | 1.7 | 1.3 | 0.9 | 1.3 | 1.3 | 0.6 |
| neutral loss | 4.3 | 2.7 | 3.2 | 16.4 | 16.0 | 13.4 |
| internal | 6.9 | 5.2 | 6.2 | 6.5 | 6.4 | 5.0 |
| immonium | 2.3 | 2.5 | 1.8 | 2.7 | 3.0 | 1.9 |
| parent | 1.1 | 1.2 | 1.6 | 2.4 | 2.3 | 2.9 |
| parent intensity | 191.6 | 204.9 | 369.1 | 197.9 | 199.1 | 295.1 |

[a] "both" means true positive for both the sequence search and the rescoring, "rescore only" means true positive only after rescoring, and "search only" means true positive only before rescoring.

---

Table 3 contains ion annotations for both phosphorylated and unphosphorylated spectra identified by Mascot searching before and after rescoring. The ion series omitted from the table showed no major changes. The values in the table are calculated only on matches to true peptides.

For unphosphorylated spectra, the "rescore" spectra have larger numbers of unannotated ions compared to the set of spectra that scored well before and after rescoring. Predicted spectra ought to work better when matching spectra with large numbers of unannotated ions since sequence searches like Mascot do not calculate unannotated ions and can assume they are mismatches, lowering scores. As noted previously, almost all existing peptide spectra prediction models do not predict this type of ion. There are also fewer y, b, neutral loss and internal ions present in the rescore only set. The specificity of sequence search scores is dependent on the number of theoretical ions considered: the more the number of theoretical ions, the more potential false positive matches. Sequence search engines attempt to be parsimonious in the number of theoretical ions considered, but these numbers still tend to be in excess of the number of ions actually in the spectrum. Library spectra do not have this issue.

The unphosphorylated spectra that are positive in the "search only" set come from longer peptides and, as shown by the maximum parent intensity, dominated by intense parent ions. This may be due to the somewhat poorer prediction performance of the network for longer peptides, as indicated in Figure S4.

Like the unphosphorylated spectra, the phosphorylated spectra that are identified only after rescoring have a larger fraction of unannotated ions and fewer b ions. The small set of phosphopeptide spectra that are identified by "search only" are longer peptides with more intense parent ions. As with the unphosphorylated peptides, this may be due to the poorer prediction performance of the model for longer peptides.

## Conclusions

In order to better identify peptides from complex biosamples that may be atypically cleaved or have multiple possible modifications, we used a deep neural network to predict complete HCD mass spectra that closely model experimental spectra. An important feature of this approach that is different from most other neural networks used to predict spectra is that predicting complete spectra does not require annotating training spectra, making it easier to train and predict modified spectra, including biologically relevant PTMs like phosphorylation. The ability of our network to generalize to multiple modifications indicates that it may be adequate to train a single large network instead of training many small networks.

Training with the NIST peptide spectral libraries generalized the network to a wide range of peptide lengths, charges, modifications, and NCE's. This includes TMT modified peptides and non-tryptically cleaved peptides. The utility of the network was further improved by algorithmically upscaling 80% of the significant ions to a resolution at the limit of calculation.

By rescoring the identifications from a sequence search over a large chemical space using spectral library search methods, we show that spectral library searching using predicted spectra should significantly outperform sequence searches. In particular, the score separation between true and false matches increases by 82%, allowing for extensive exploration of chemical space. Matching predicted spectra that consist solely of annotated ions yields a score separation of 24%, revealing that most of the improvement in score separation is due to unannotated ions. Sequence searches and most peptide spectral prediction networks do not calculate these unannotated ions.

Our test set consists of high quality library spectra previously identified via sequence searches and whose peptides are sequence dissimilar to the training set, making improvements in identification difficult. However, by rescoring the sequence search of our test set, we show a net 8% increase in identifications of all peptides, a 21% increase in non-tryptic peptides and a 17% increase in identified phosphopeptides at 1% FDR. An analysis of the improved identifications indicates that the increase for phosphorylated and unphosphorylated peptides comes in large part from matching unannotated ions, not from canonical ion series. The increase for unphosphorylated peptides also appears due to fewer false positive matches to theoretical ions not actually present in the query or library spectra. These results are consistent with the general architecture of sequence search algorithms, as they match to theoretical spectra of typically unit intensity that rely entirely on the presence or absence of backbone cleavages at standard sites, while spectral searches focus on the full range intensities of any possible ion type. Predicting the complete spectra of modified and unmodified peptides, including ion types due to unknown or unannotated mechanisms, results in substantial improvements in the identification of peptide spectra in a large chemical space.

The main obstacle to the use of spectral libraries in proteomics has been incomplete coverage of the proteome, particularly for modifications. We have found that the use of AI predictions to supplement experimental spectral libraries, which we denote as "Alomics", may overcome the barrier of limited coverage of the proteome by not only searching larger areas of the proteome but also by significantly improving the ability to discriminate between true and false hits. While these supplemented libraries should be useful in improving identifications from more extensive sequence searches and screening false positives from open searches, these results indicate that the benefits of directly searching mass spectral libraries can now be realized generally for deep exploration of the proteome.

For applications where the prediction uncertainty and set of modifications supported by the network are sufficient for regular use, it is possible to generate libraries that consist entirely of predicted spectra as there is no requirement to have experimental spectra when doing the predictions. We can conclude from our results that it will help in identification via searching, but whether this predicted-only approach works for particular applications such as Data Independent Acquisition (DIA) remains to be tested, but is a question of interest as predicting full spectra may allow the deconvolution of mixture spectra without requiring an experimental spectral library.

A specific application that may benefit from our method is the detection of the spectra of naturally occurring peptides, including immunopeptide spectra. Immunopeptides, especially MHC class I, are short, nonspecifically cleaved peptides that can originate from pathogens or neoplasms. Because the search space of immunopeptide spectra is very large and the number of ions in each spectra is relatively small, identification of these spectra is challenging. As described in the Results, the prediction performance of our network on non-tryptic peptide spectra approaches the performance on tryptic peptide spectra, making it likely that our method can help improve the detection of naturally occurring peptides.

There are multiple pragmatic benefits to the approach taken in this study: A single DNN can be trained using unannotated spectra to predict the complete spectra of peptides with many different modifications and types of cleavage. This DNN can be used centrally to augment experimental spectral libraries, eliminating any need for the special hardware and software typically required to run AI networks quickly. Finally, these augmented libraries can be searched by existing spectral library search algorithms, including those that demonstrate high performance[17,62].

# Author Information

## Corresponding Author

**Lewis Y. Geer** --- *Mass Spectrometry Data Center, Biomolecular Measurement Division, National Institute of Standards and Technology, 100 Bureau Dr., Gaithersburg, Maryland 20899, United States*; Phone: +1 301 975 6820; Email: lewis.geer@nist.gov

## Authors

**Joel Lapin** --- *Department of Physics, Georgetown University, Washington, DC 20057; Associate, Mass Spectrometry Data Center, Biomolecular Measurement Division, National Institute of Standards and Technology, 100 Bureau Dr., Gaithersburg, Maryland 20899, United States*

**Douglas J. Slotta** --- *Mass Spectrometry Data Center, Biomolecular Measurement Division, National Institute of Standards and Technology, 100 Bureau Dr., Gaithersburg, Maryland 20899, United States*

**Tytus D. Mak** --- *Mass Spectrometry Data Center, Biomolecular Measurement Division, National Institute of Standards and Technology, 100 Bureau Dr., Gaithersburg, Maryland 20899, United States*

**Stephen E. Stein** --- *Mass Spectrometry Data Center, Biomolecular Measurement Division, National Institute of Standards and Technology, 100 Bureau Dr., Gaithersburg, Maryland 20899, United States*

## Notes

The authors declare no competing financial interests. All commercial instruments, software, and materials used in the study are for experimental purposes only. Such identification does not intend recommendation or endorsement by the National Institute of Standards and Technology, nor does it intend that the materials, software, or instruments used are necessarily the best available for the purpose.

# Supporting Information

The following supporting information is available free of charge at ACS website http://pubs.acs.org

Figure S1. Stein-Scott correction.

Figures S2-S4. Variation of the similarity score S over features of the test set.

Figure S5. Score histograms for predicted spectra containing only annotated ions.

(PDF)

## Acknowledgements

We are grateful to the NIST FARSAIT program for their support. We thank Dr. William E. Wallace for many useful comments and Dr. Sergey L. Sheetlin for helping us use the NIST peptide spectral library. The authors also appreciate the efforts of the reviewers in improving this paper. This research was conducted in part through the support of a NIST cooperative agreement with Georgetown University. This article is dedicated to the memory of Sanford P. Markey.

## References

(1)  Eng, J. K.; Jahan, T. A.; Hoopmann, M. R. Comet: An Open-Source MS/MS Sequence Database Search Tool. *PROTEOMICS* **2013**, *13* (1), 22–24. https://doi.org/10.1002/pmic.201200439.

(2)  Perkins, D. N.; Pappin, D. J. C.; Creasy, D. M.; Cottrell, J. S. Probability-Based Protein Identification by Searching Sequence Databases Using Mass Spectrometry Data. *Electrophoresis* **1999**, *20* (18), 3551–3567. https://doi.org/10.1002/(SICI)1522-2683(19991201)20:18<3551::AID-ELPS3551>3.0.CO;2-2.

(3)  Shilov, I. V.; Seymour, S. L.; Patel, A. A.; Loboda, A.; Tang, W. H.; Keating, S. P.; Hunter, C. L.; Nuwaysir, L. M.; Schaeffer, D. A. The Paragon Algorithm, a Next Generation Search Engine That Uses Sequence Temperature Values and Feature Probabilities to Identify Peptides from Tandem Mass Spectra. *Mol. Cell. Proteomics* **2007**, *6* (9), 1638–1655. https://doi.org/10.1074/mcp.T600050-MCP200.

(4)  Eng, J. K.; McCormack, A. L.; Yates, J. R. An Approach to Correlate Tandem Mass Spectral Data of Peptides with Amino Acid Sequences in a Protein Database. *J. Am. Soc. Mass Spectrom.* **1994**, *5* (11), 976–989. https://doi.org/10.1016/1044-0305(94)80016-2.

(5)  Craig, R.; Beavis, R. C. TANDEM: Matching Proteins with Tandem Mass Spectra. *Bioinformatics* **2004**, *20* (9), 1466–1467. https://doi.org/10.1093/bioinformatics/bth092.

(6)  Kim, S.; Pevzner, P. A. MS-GF+ Makes Progress towards a Universal Database Search Tool for Proteomics. *Nat. Commun.* **2014**, *5* (1), 5277. https://doi.org/10.1038/ncomms6277.

(7)  Cox, J.; Mann, M. MaxQuant Enables High Peptide Identification Rates, Individualized p.p.b.-Range Mass Accuracies and Proteome-Wide Protein Quantification. *Nat. Biotechnol.* **2008**, *26* (12), 1367–1372. https://doi.org/10.1038/nbt.1511.

(8)  Tu, C.; Sheng, Q.; Li, J.; Ma, D.; Shen, X.; Wang, X.; Shyr, Y.; Yi, Z.; Qu, J. Optimization of Search Engines and Postprocessing Approaches to Maximize

Peptide and Protein Identification for High-Resolution Mass Data. *J. Proteome Res.* **2015**, *14* (11), 4662–4673. https://doi.org/10.1021/acs.jproteome.5b00536.

(9)  Tang, W. H.; Halpern, B. R.; Shilov, I. V.; Seymour, S. L.; Keating, S. P.; Loboda, A.; Patel, A. A.; Schaeffer, D. A.; Nuwaysir, L. M. Discovering Known and Unanticipated Protein Modifications Using MS/MS Database Searching. *Anal. Chem.* **2005**, *77* (13), 3931–3946. https://doi.org/10.1021/ac0481046.

(10) Solntsev, S. K.; Shortreed, M. R.; Frey, B. L.; Smith, L. M. Enhanced Global Post-Translational Modification Discovery with MetaMorpheus. *J. Proteome Res.* **2018**, *17* (5), 1844–1851. https://doi.org/10.1021/acs.jproteome.7b00873.

(11) Na, S.; Bandeira, N.; Paek, E. Fast Multi-Blind Modification Search through Tandem Mass Spectrometry. *Mol. Cell. Proteomics* **2012**, *11* (4), M111.010199. https://doi.org/10.1074/mcp.M111.010199.

(12) Bittremieux, W.; Laukens, K.; Noble, W. S. Extremely Fast and Accurate Open Modification Spectral Library Searching of High-Resolution Mass Spectra Using Feature Hashing and Graphics Processing Units. *J. Proteome Res.* **2019**, *18* (10), 3792–3799. https://doi.org/10.1021/acs.jproteome.9b00291.

(13) Yu, F.; Teo, G. C.; Kong, A. T.; Haynes, S. E.; Avtonomov, D. M.; Geiszler, D. J.; Nesvizhskii, A. I. Identification of Modified Peptides Using Localization-Aware Open Search. *Nat. Commun.* **2020**, *11* (1), 4065. https://doi.org/10.1038/s41467-020-17921-y.

(14) Yu, F.; Li, N.; Yu, W. PIPI: PTM-Invariant Peptide Identification Using Coding Method. *J. Proteome Res.* **2016**, *15* (12), 4423–4435. https://doi.org/10.1021/acs.jproteome.6b00485.

(15) Devabhaktuni, A.; Lin, S.; Zhang, L.; Swaminathan, K.; Gonzales, C.; Olsson, N.; Pearlman, S.; Rawson, K.; Elias, J. E. TagGraph Reveals Vast Protein Modification Landscapes from Large Tandem Mass Spectrometry Data Sets. *Nat. Biotechnol.* **2019**, *37* (4), 469–479. https://doi.org/10.1038/s41587-019-0067-5.

(16) Matthiesen, R. Solution to Dark Matter Identified by Mass-Tolerant Database Search. In *Mass Spectrometry Data Analysis in Proteomics*; Matthiesen, R., Ed.; Methods in Molecular Biology; Springer: New York, NY, 2020; pp 231–240. https://doi.org/10.1007/978-1-4939-9744-2_9.

(17) Lam, H.; Deutsch, E. W.; Eddes, J. S.; Eng, J. K.; King, N.; Stein, S. E.; Aebersold, R. Development and Validation of a Spectral Library Searching Method for Peptide Identification from MS/MS. *Proteomics* **2007**, *7* (5), 655–667. https://doi.org/10.1002/pmic.200600625.

(18) Deutsch, E. W.; Perez-Riverol, Y.; Chalkley, R. J.; Wilhelm, M.; Tate, S.; Sachsenberg, T.; Walzer, M.; Käll, L.; Delanghe, B.; Böcker, S.; Schymanski, E. L.; Wilmes, P.; Dorfer, V.; Kuster, B.; Volders, P.-J.; Jehmlich, N.; Vissers, J. P. C.; Wolan, D. W.; Wang, A. Y.; Mendoza, L.; Shofstahl, J.; Dowsey, A. W.; Griss, J.; Salek, R. M.; Neumann, S.; Binz, P.-A.; Lam, H.; Vizcaíno, J. A.; Bandeira, N.; Röst, H. Expanding the Use of Spectral Libraries in Proteomics. *J. Proteome Res.* **2018**, *17* (12), 4051–4060. https://doi.org/10.1021/acs.jproteome.8b00485.

(19) Burke, M. C.; Mirokhin, Y. A.; Tchekhovskoi, D. V.; Markey, S. P.; Heidbrink Thompson, J.; Larkin, C.; Stein, S. E. The Hybrid Search: A Mass Spectral Library Search Method for Discovery of Modifications in Proteomics. *J. Proteome Res.* **2017**, *16* (5), 1924–1935. https://doi.org/10.1021/acs.jproteome.6b00988.

(20) Shiferaw, G. A.; Vandermarliere, E.; Hulstaert, N.; Gabriels, R.; Martens, L.; Volders, P.-J. COSS: A Fast and User-Friendly Tool for Spectral Library Searching. *J. Proteome Res.* **2020**, *19* (7), 2786–2793. https://doi.org/10.1021/acs.jproteome.9b00743.

(21) Zhang, X.; Li, Y.; Shao, W.; Lam, H. Understanding the Improved Sensitivity of Spectral Library Searching over Sequence Database Searching in Proteomics Data Analysis. *PROTEOMICS* **2011**, *11* (6), 1075–1085. https://doi.org/10.1002/pmic.201000492.

(22) Schütz, F.; Kapp, E. A.; Simpson, R. J.; Speed, T. P. Deriving Statistical Models for Predicting Peptide Tandem MS Product Ion Intensities. *Biochem. Soc. Trans.* **2003**, *31* (6), 1479–1483. https://doi.org/10.1042/bst0311479.

(23) Li, W.; Ji, L.; Goya, J.; Tan, G.; Wysocki, V. H. SQID: An Intensity-Incorporated Protein Identification Algorithm for Tandem Mass Spectrometry. *J. Proteome Res.* **2011**, *10* (4), 1593–1602. https://doi.org/10.1021/pr100959y.

(24) Zhang, Z. Prediction of Low-Energy Collision-Induced Dissociation Spectra of Peptides. *Anal. Chem.* **2004**, *76* (14), 3908–3922. https://doi.org/10.1021/ac049951b.

(25) Zhang, Z. Prediction of Low-Energy Collision-Induced Dissociation Spectra of Peptides with Three or More Charges. *Anal. Chem.* **2005**, *77* (19), 6364–6373. https://doi.org/10.1021/ac050857k.

(26) Tiwary, S.; Levy, R.; Gutenbrunner, P.; Salinas Soto, F.; Palaniappan, K. K.; Deming, L.; Berndl, M.; Brant, A.; Cimermancic, P.; Cox, J. High-Quality MS/MS Spectrum Prediction for Data-Dependent and Data-Independent Acquisition Data Analysis. *Nat. Methods* **2019**, *16* (6). https://doi.org/10.1038/s41592-019-0427-6.

(27) Yang, Y.; Liu, X.; Shen, C.; Lin, Y.; Yang, P.; Qiao, L. In Silico Spectral Libraries by Deep Learning Facilitate Data-Independent Acquisition Proteomics. *Nat. Commun.* **2020**, *11* (1). https://doi.org/10.1038/s41467-019-13866-z.

(28) Lin, Y.-M.; Chen, C.-T.; Chang, J.-M. MS2CNN: Predicting MS/MS Spectrum Based on Protein Sequence Using Deep Convolutional Neural Networks. *BMC Genomics* **2019**, *20* (S9), 906. https://doi.org/10.1186/s12864-019-6297-6.

(29) Degroeve, S.; Martens, L. MS2PIP: A Tool for MS/MS Peak Intensity Prediction. *Bioinformatics* **2013**, *29* (24). https://doi.org/10.1093/bioinformatics/btt544.

(30) Guan, S.; Moran, M. F.; Ma, B. Prediction of LC-MS/MS Properties of Peptides from Sequence by Deep Learning. *Mol. Cell. Proteomics* **2019**, *18* (10). https://doi.org/10.1074/mcp.TIR119.001412.

(31) Elias, J. E.; Gibbons, F. D.; King, O. D.; Roth, F. P.; Gygi, S. P. Intensity-Based Protein Identification by Machine Learning from a Library of Tandem Mass Spectra. *Nat. Biotechnol.* **2004**, *22* (2). https://doi.org/10.1038/nbt930.

(32) Gessulat, S.; Schmidt, T.; Zolg, D. P.; Samaras, P.; Schnatbaum, K.; Zerweck, J.; Knaute, T.; Rechenberger, J.; Delanghe, B.; Huhmer, A.; Reimer, U.; Ehrlich, H.-C.; Aiche, S.; Kuster, B.; Wilhelm, M. Prosit: Proteome-Wide Prediction of Peptide Tandem Mass Spectra by Deep Learning. *Nat. Methods* **2019**, *16* (6). https://doi.org/10.1038/s41592-019-0426-7.

(33) Ekvall, M.; Truong, P.; Gabriel, W.; Wilhelm, M.; Käll, L. Prosit Transformer: A Transformer for Prediction of MS2 Spectrum Intensities. *J. Proteome Res.* **2022**, *21* (5), 1359–1364. https://doi.org/10.1021/acs.jproteome.1c00870.

(34) Zhou, X.-X.; Zeng, W.-F.; Chi, H.; Luo, C.; Liu, C.; Zhan, J.; He, S.-M.; Zhang, Z. PDeep: Predicting MS/MS Spectra of Peptides with Deep Learning. *Anal. Chem.* **2017**, *89* (23), 12690–12697. https://doi.org/10.1021/acs.analchem.7b02566.

(35) Zeng, W.-F.; Zhou, X.-X.; Zhou, W.-J.; Chi, H.; Zhan, J.; He, S.-M. MS/MS Spectrum Prediction for Modified Peptides Using PDeep2 Trained by Transfer Learning. *Anal. Chem.* **2019**, *91* (15), 9724–9731. https://doi.org/10.1021/acs.analchem.9b01262.

(36) Tarn, C.; Zeng, W.-F. PDeep3: Toward More Accurate Spectrum Prediction with Fast Few-Shot Learning. *Anal. Chem.* **2021**. https://doi.org/10.1021/acs.analchem.0c05427.

(37) Chalmers, M. J.; Kolch, W.; Emmett, M. R.; Marshall, A. G.; Mischak, H. Identification and Analysis of Phosphopeptides. *J. Chromatogr. B Analyt. Technol. Biomed. Life. Sci.* **2004**, *803* (1), 111–120. https://doi.org/10.1016/j.jchromb.2003.09.006.

(38) Quadroni, M.; James, P. Phosphopeptide Analysis. In *Proteomics in Functional Genomics: Protein Structure Analysis*; Jollès, P., Jörnvall, H., Eds.; Birkhäuser: Basel, 2000; pp 199–213. https://doi.org/10.1007/978-3-0348-8458-7_13.

(39) Liu, K.; Li, S.; Wang, L.; Ye, Y.; Tang, H. Full-Spectrum Prediction of Peptides Tandem Mass Spectra Using Deep Neural Network. *Anal. Chem.* **2020**, *92* (6). https://doi.org/10.1021/acs.analchem.9b04867.

(40) Lapin, J.; Dong, Q.; Mak, T.; Slotta, D.; Geer, L. Characterizing the Out-of-Distribution Behavior of a Deep Machine Learning Model for Proteomics MS/MS Spectra Prediction; 70th ASMS Conference on Mass Spectrometry and Allied Topics, 2022.

(41) Lapin, J.; Dong, Q.; Mak, T.; Slotta, D.; Geer, L. Uncertainty Quantification for Prediction of MS/MS Spectra via Deep Learning; 69th ASMS Conference on Mass Spectrometry and Allied Topics, 2021.

(42) Sheetlin, S.; Wang, G.; Tchekhovskoi, D.; Zhang, Z.; Stein, S. Filtering and Optimization of Peptide Tandem Mass Spectral Libraries; 68th ASMS Conference on Mass Spectrometry and Allied Topics, 2020.

(43) Stein, S. Mass Spectral Reference Libraries: An Ever-Expanding Resource for Chemical Identification. *Anal. Chem.* **2012**, *84* (17), 7274–7282. https://doi.org/10.1021/ac301205z.

(44) Michalski, A.; Neuhauser, N.; Cox, J.; Mann, M. A Systematic Investigation into the Nature of Tryptic HCD Spectra. *J. Proteome Res.* **2012**, *11* (11), 5479–5491. https://doi.org/10.1021/pr3007045.

(45) Zolg, D. P.; Gessulat, S.; Paschke, C.; Graber, M.; Rathke‐Kuhnert, M.; Seefried, F.; Fitzemeier, K.; Berg, F.; Lopez‐Ferrer, D.; Horn, D.; Henrich, C.; Huhmer, A.; Delanghe, B.; Frejno, M. INFERYS Rescoring: Boosting Peptide Identifications and Scoring Confidence of Database Search Results. *Rapid Commun. Mass Spectrom.* **2021**. https://doi.org/10.1002/rcm.9128.

(46) Gabriel, W.; The, M.; Zolg, D. P.; Bayer, F. P.; Shouman, O.; Lautenbacher, L.; Schnatbaum, K.; Zerweck, J.; Knaute, T.; Delanghe, B.; Huhmer, A.; Wenschuh, H.; Reimer, U.; Médard, G.; Kuster, B.; Wilhelm, M. Prosit-TMT: Deep Learning Boosts Identification of TMT-Labeled Peptides. *Anal. Chem.* **2022**, *94* (20), 7181–7190. https://doi.org/10.1021/acs.analchem.1c05435.

(47) Wilhelm, M.; Zolg, D. P.; Graber, M.; Gessulat, S.; Schmidt, T.; Schnatbaum, K.; Schwencke-Westphal, C.; Seifert, P.; de Andrade Krätzig, N.; Zerweck, J.; Knaute, T.; Bräunlein, E.; Samaras, P.; Lautenbacher, L.; Klaeger, S.; Wenschuh, H.; Rad, R.; Delanghe, B.; Huhmer, A.; Carr, S. A.; Clauser, K. R.; Krackhardt, A. M.; Reimer, U.; Kuster, B. Deep Learning Boosts Sensitivity of Mass Spectrometry-Based Immunopeptidomics. *Nat. Commun.* **2021**, *12* (1), 3346. https://doi.org/10.1038/s41467-021-23713-9.

(48) Na, S.; Choi, H.; Paek, E. Deephos: Predicted Spectral Database Search for TMT-Labeled Phosphopeptides and Its False Discovery Rate Estimation. *Bioinformatics* **2022**, *38* (11), 2980–2987. https://doi.org/10.1093/bioinformatics/btac280.

(49) *Libraries of Peptide Tandem Mass Spectra.* https://chemdata.nist.gov/dokuwiki/doku.php?id=peptidew:cdownload (accessed 2020-07-20).

(50) Dong, Q.; Liang, Y.; Yan, X.; Markey, S. P.; Mirokhin, Y. A.; Tchekhovskoi, D. V.; Bukhari, T. H.; Stein, S. E. The NISTmAb Tryptic Peptide Spectral Library for Monoclonal Antibody Characterization. *mAbs* **2018**, *10* (3), 354–369. https://doi.org/10.1080/19420862.2018.1436921.

(51) Sharma, K.; D'Souza, R. C. J.; Tyanova, S.; Schaab, C.; Wiśniewski, J. R.; Cox, J.; Mann, M. Ultradeep Human Phosphoproteome Reveals a Distinct Regulatory Nature of Tyr and Ser/Thr-Based Signaling. *Cell Rep.* **2014**, *8* (5), 1583–1594. https://doi.org/10.1016/j.celrep.2014.07.036.

(52) Paszke, A.; Gross, S.; Massa, F.; Lerer, A.; Bradbury, J.; Chanan, G.; Killeen, T.; Lin, Z.; Gimelshein, N.; Antiga, L.; Desmaison, A.; Kopf, A.; Yang, E.; DeVito, Z.; Raison, M.; Tejani, A.; Chilamkurthy, S.; Steiner, B.; Fang, L.; Bai, J.; Chintala, S. PyTorch: An Imperative Style, High-Performance Deep Learning Library. **2019**. https://doi.org/10.48550/arXiv.1912.01703.

(53) Ranganath Krishnan; Pi Esposito; Subedar, M. Bayesian-Torch: Bayesian Neural Network Layers for Uncertainty Estimation, 2022. https://doi.org/10.5281/ZENODO.5908307.

(54) Falcon, W.; Borovec, J.; Wälchli, A.; Eggert, N.; Schock, J.; Jordan, J.; Skafte, N.; Ir1dXD; Bereznyuk, V.; Harris, E.; Tullie Murrell; Yu, P.; Præsius, S.; Addair, T.; Zhong, J.; Lipin, D.; Uchida, S.; Shreyas Bapat; Schröter, H.; Dayma, B.; Karnachev, A.; Akshay Kulkarni; Shunta Komatsu; Martin.B; Jean-Baptiste SCHIRATTI; Mary, H.; Byrne, D.; Cristobal Eyzaguirre; Cinjon; Bakhtin, A. PyTorchLightning/Pytorch-Lightning: 0.7.6 Release, 2020. https://doi.org/10.5281/ZENODO.3828935.

(55) Virtanen, P.; Gommers, R.; Oliphant, T. E.; Haberland, M.; Reddy, T.; Cournapeau, D.; Burovski, E.; Peterson, P.; Weckesser, W.; Bright, J.; van der Walt, S. J.; Brett, M.; Wilson, J.; Millman, K. J.; Mayorov, N.; Nelson, A. R. J.; Jones, E.; Kern, R.; Larson, E.; Carey, C. J.; Polat, İ.; Feng, Y.; Moore, E. W.; VanderPlas, J.; Laxalde, D.; Perktold, J.; Cimrman, R.; Henriksen, I.; Quintero, E. A.; Harris, C. R.; Archibald, A. M.; Ribeiro, A. H.; Pedregosa, F.; van Mulbregt, P. SciPy 1.0: Fundamental Algorithms for Scientific Computing in Python. *Nat. Methods* **2020**, *17* (3), 261–272. https://doi.org/10.1038/s41592-019-0686-2.

(56) McKinney, W. Data Structures for Statistical Computing in Python; Austin, Texas,

2010; pp 56–61. https://doi.org/10.25080/Majora-92bf1922-00a.

(57) Harris, C. R.; Millman, K. J.; van der Walt, S. J.; Gommers, R.; Virtanen, P.; Cournapeau, D.; Wieser, E.; Taylor, J.; Berg, S.; Smith, N. J.; Kern, R.; Picus, M.; Hoyer, S.; van Kerkwijk, M. H.; Brett, M.; Haldane, A.; del Río, J. F.; Wiebe, M.; Peterson, P.; Gérard-Marchant, P.; Sheppard, K.; Reddy, T.; Weckesser, W.; Abbasi, H.; Gohlke, C.; Oliphant, T. E. Array Programming with NumPy. *Nature* **2020**, *585* (7825), 357–362. https://doi.org/10.1038/s41586-020-2649-2.

(58) Stein, S. E.; Scott, D. R. Optimization and Testing of Mass Spectral Library Search Algorithms for Compound Identification. *J. Am. Soc. Mass Spectrom.* **1994**, *5* (9), 859–866. https://doi.org/10.1016/1044-0305(94)87009-8.

(59) The UniProt Consortium; Bateman, A.; Martin, M.-J.; Orchard, S.; Magrane, M.; Agivetova, R.; Ahmad, S.; Alpi, E.; Bowler-Barnett, E. H.; Britto, R.; Bursteinas, B.; Bye-A-Jee, H.; Coetzee, R.; Cukura, A.; Da Silva, A.; Denny, P.; Dogan, T.; Ebenezer, T.; Fan, J.; Castro, L. G.; Garmiri, P.; Georghiou, G.; Gonzales, L.; Hatton-Ellis, E.; Hussein, A.; Ignatchenko, A.; Insana, G.; Ishtiaq, R.; Jokinen, P.; Joshi, V.; Jyothi, D.; Lock, A.; Lopez, R.; Luciani, A.; Luo, J.; Lussi, Y.; MacDougall, A.; Madeira, F.; Mahmoudy, M.; Menchi, M.; Mishra, A.; Moulang, K.; Nightingale, A.; Oliveira, C. S.; Pundir, S.; Qi, G.; Raj, S.; Rice, D.; Lopez, M. R.; Saidi, R.; Sampson, J.; Sawford, T.; Speretta, E.; Turner, E.; Tyagi, N.; Vasudev, P.; Volynkin, V.; Warner, K.; Watkins, X.; Zaru, R.; Zellner, H.; Bridge, A.; Poux, S.; Redaschi, N.; Aimo, L.; Argoud-Puy, G.; Auchincloss, A.; Axelsen, K.; Bansal, P.; Baratin, D.; Blatter, M.-C.; Bolleman, J.; Boutet, E.; Breuza, L.; Casals-Casas, C.; de Castro, E.; Echioukh, K. C.; Coudert, E.; Cuche, B.; Doche, M.; Dornevil, D.; Estreicher, A.; Famiglietti, M. L.; Feuermann, M.; Gasteiger, E.; Gehant, S.; Gerritsen, V.; Gos, A.; Gruaz-Gumowski, N.; Hinz, U.; Hulo, C.; Hyka-Nouspikel, N.; Jungo, F.; Keller, G.; Kerhornou, A.; Lara, V.; Le Mercier, P.; Lieberherr, D.; Lombardot, T.; Martin, X.; Masson, P.; Morgat, A.; Neto, T. B.; Paesano, S.; Pedruzzi, I.; Pilbout, S.; Pourcel, L.; Pozzato, M.; Pruess, M.; Rivoire, C.; Sigrist, C.; Sonesson, K.; Stutz, A.; Sundaram, S.; Tognolli, M.; Verbregue, L.; Wu, C. H.; Arighi, C. N.; Arminski, L.; Chen, C.; Chen, Y.; Garavelli, J. S.; Huang, H.; Laiho, K.; McGarvey, P.; Natale, D. A.; Ross, K.; Vinayaka, C. R.; Wang, Q.; Wang, Y.; Yeh, L.-S.; Zhang, J.; Ruch, P.; Teodoro, D. UniProt: The Universal Protein Knowledgebase in 2021. *Nucleic Acids Res.* **2021**, *49* (D1), D480–D489. https://doi.org/10.1093/nar/gkaa1100.

(60) Sturm, M.; Bertsch, A.; Gröpl, C.; Hildebrandt, A.; Hussong, R.; Lange, E.; Pfeifer, N.; Schulz-Trieglaff, O.; Zerck, A.; Reinert, K.; Kohlbacher, O. OpenMS – An Open-Source Software Framework for Mass Spectrometry. *BMC Bioinformatics* **2008**, *9* (1), 163. https://doi.org/10.1186/1471-2105-9-163.

(61) Elias, J. E.; Gygi, S. P. Target-Decoy Search Strategy for Increased Confidence in Large-Scale Protein Identifications by Mass Spectrometry. *Nat. Methods* **2007**, *4* (3), 207–214. https://doi.org/10.1038/nmeth1019.

(62) Wang, L.; Liu, K.; Li, S.; Tang, H. A Fast and Memory-Efficient Spectral Library Search Algorithm Using Locality-Sensitive Hashing. *Proteomics* **2020**, *20* (21–22), e2000002. https://doi.org/10.1002/pmic.202000002.

# AIomics: exploring more of the proteome using mass spectral libraries extended by AI: Supporting Information

Figure S1. Stein-Scott correction.
Figures S2-S4. Variation of the similarity score $S$ over features of the test set.
Figure S5. Score histograms for predicted spectra containing only annotated ions.

# Correction to the Stein-Scott Dot Product

A known issue with the Stein-Scott dot product $S$ is that the significance threshold becomes higher for spectra with very few ions. The reason for this can be understood using a geometrical argument: the size of the search space for a dot product is proportional to the number of dimensions, and since the number of dimensions is low for spectra with few ions, there are more chances to make an incorrect match. While there is no closed form expression for this threshold, we can use the decoy matches to fit an exponential approximation to the empirical threshold. In the graph below, the plotted points are the 90% highest cosine score of predicted decoy spectra matched to the query spectra from our validation set, where the x axis is the number of significant ions in the query spectra.

A non-linear least squares fit of an exponential using scipy curve_fit yields the correction given in the legend of the graph. The exponential term of this correction is subtracted from the Stein-Scott dot product to give the corrected similarity score $S$. This subtraction may result in negative values to indicate lack of significance. If negative numbers are undesirable, the score can be clipped at 0 without harm.

## Figure S1. Stein-Scott Correction

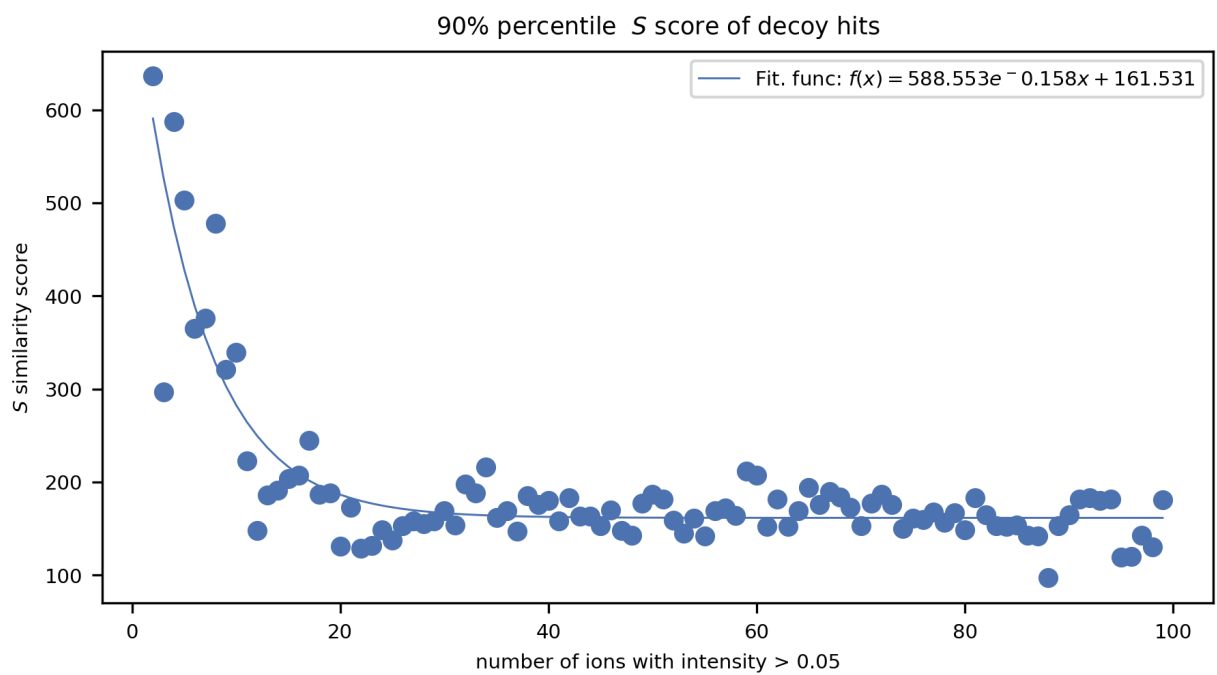

Figure S1. Plot of the 90th percentile Stein-Scott dot product of predicted decoy spectra to experimental spectra from the validation set with a given number of higher intensity ions per spectrum. An exponential function, given in the legend, is fit to the data.

The corrected Stein-Scott dot product is not normalized to the number of library spectra with matching numbers of high intensity ions. Examining how the correction is created makes this clear: for a given number of high intensity ions, we take all decoy spectra with that number of high intensity ions and take the 90% percentile of the uncorrected Stein-Scott score. Taking the 90% percentile instead of the top hit makes the correction independent of the number of peptides in the search library that have the same number of high intensity ions. This is because the underlying distribution being sampled is independent of the number of matching spectra.

## Variation of the similarity score *S* over features of the test set

For the features of NCE, precursor charge, and peptide length, we computed hexagonal binning plots of the number of test spectra and their similarity scores. These plots indicate how well the network we are using predicts spectra over the entire feature range of the test set. No attempt has been made to decorrelate these features to understand their individual, uncorrelated contribution to the similarity score, but these plots indicate that the model is useful over a range of feature values typically found in proteomics experiments.

Figure S2.

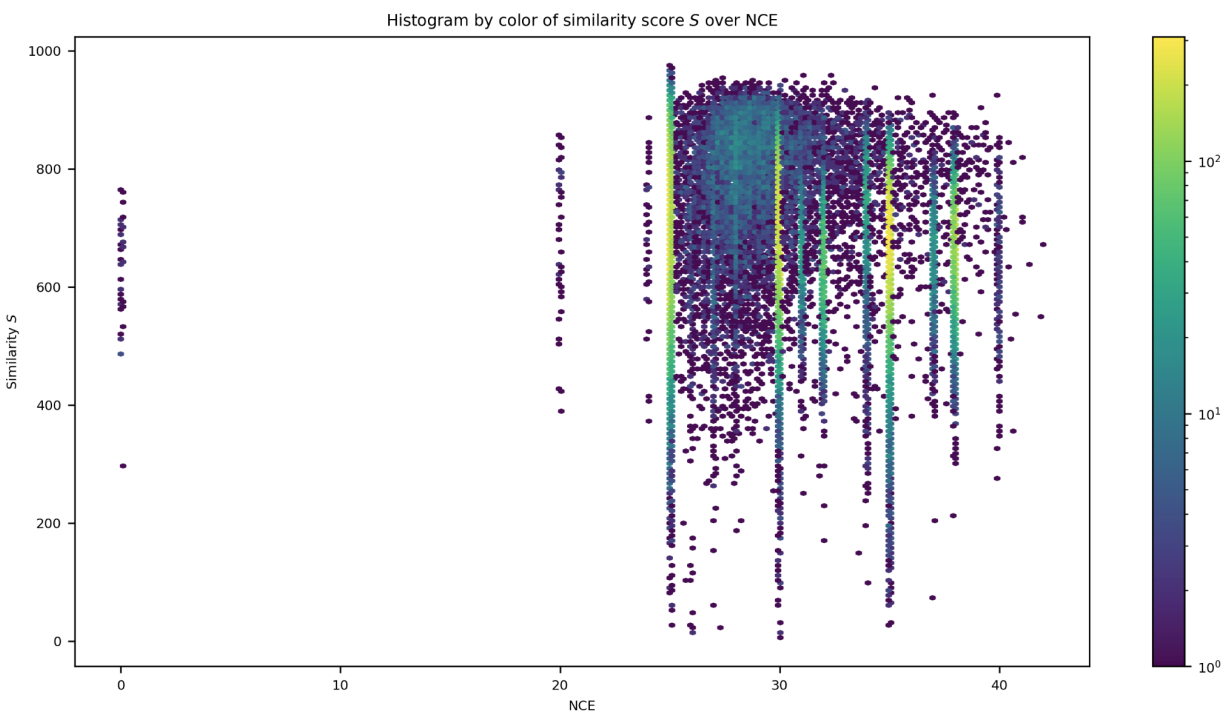

Figure S2. 2D hexagonal histogram of spectra from the test set, binned by the Stein-Scott dot product of the experimental spectra to their predicted spectra and NCE. Counts of spectra are indicated by the color legend to the right.

Figure S3.

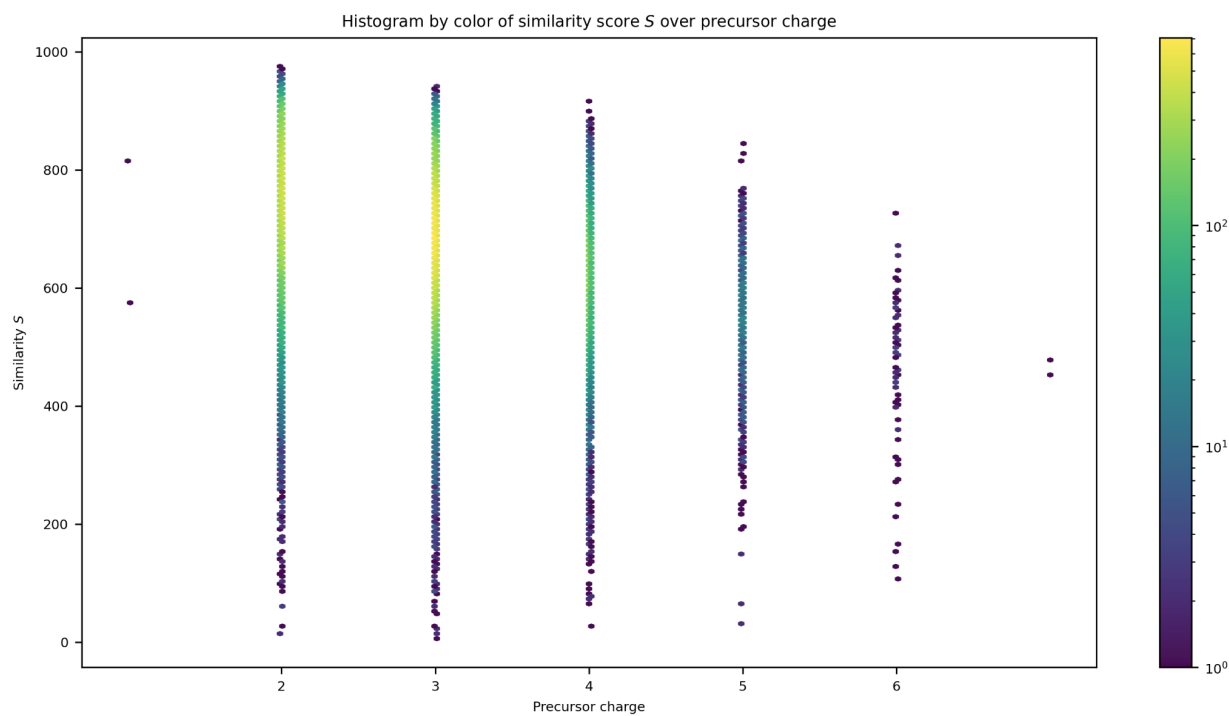

Figure S3. 2D hexagonal histogram of spectra from the test set, binned by the Stein-Scott dot product of the experimental spectra to their predicted spectra and precursor charge. Counts of spectra are indicated by the color legend to the right.

Figure S4.

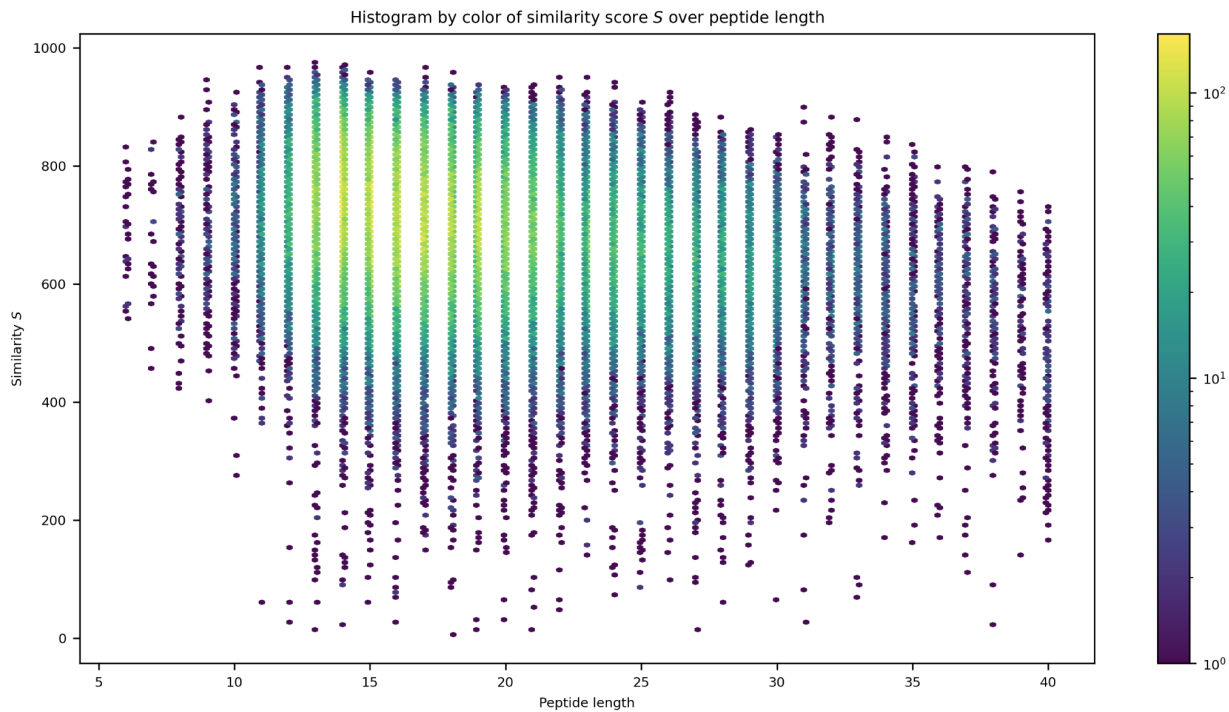

Figure S4. 2D hexagonal histogram of spectra from the test set, binned by the Stein-Scott dot product of the experimental spectra to their predicted spectra and peptide length. Counts of spectra are indicated by the color legend to the right.

# Figure S5. Score histograms for predicted spectra containing only annotated ions

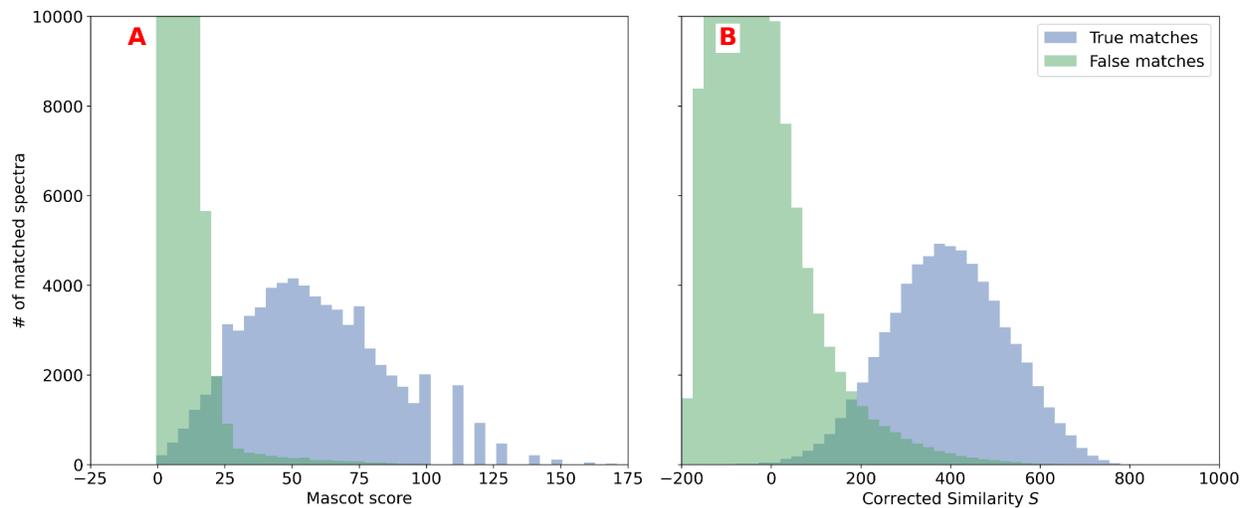

Figure S5. (A) Histogram of the Mascot ions score for both true and false matches to the test set spectra, searching against the human, mouse, and Chinese hamster proteome. (B) histogram of the corrected $S$ score as applied to predicted spectra that contain only annotated ions.